\documentclass[%
 reprint,
 amsmath,amssymb,
 aps,pra
]{revtex4-2}

\usepackage[table]{xcolor}

\usepackage{siunitx}
\usepackage{graphicx}
\usepackage{dcolumn}
\usepackage{bm}
\usepackage[unicode=true, colorlinks=true, citecolor={blue!80!black}, urlcolor={blue!50!black}, linkcolor = {blue!80!black}]{hyperref}
\usepackage{easyReview}
\usepackage{adjustbox}
\usepackage{xcolor}
\usepackage[T1]{fontenc}
\usepackage{blindtext}
\usepackage{braket}

\begin{document}

\preprint{QuTech/AndersenLab}

\title{Gate-tunable phase transition in a bosonic Su-Schrieffer-Heeger chain}

\author{Lukas Johannes Splitthoff$^{1,2}$}
\email{l.j.splitthoff@gmail.com}
\author{Miguel Carrera Belo$^{1,2}$}
\author{Guliuxin Jin$^{2}$}
\author{Yu Li$^{3}$}
\author{Eliska Greplova$^{2}$}
\author{Christian Kraglund Andersen$^{1,2}$}

\affiliation{$^1$QuTech, Delft University of Technology, Delft 2628 CJ, The Netherlands}
\affiliation{$^2$Kavli Institute for Nanoscience, Delft University of Technology, Delft 2628 CJ, The Netherlands}
\affiliation{$^3$Center for Quantum Devices, Niels Bohr Institute, University of Copenhagen, 2100 Copenhagen, Denmark}

\date{\today}

\begin{abstract}
    Metamaterials engineered to host topological states of matter in controllable quantum systems hold promise for the advancement of quantum simulations and quantum computing technologies. 
    In this context, the Su-Schrieffer-Heeger (SSH) model has gained prominence due to its simplicity and practical applications. Here, we present the implementation of a gate-tunable, five-unit-cell bosonic SSH chain on a one-dimensional lattice of superconducting resonators. We achieve electrostatic control over the inductive intra-cell coupling using semiconductor nanowire junctions, which enables the spectroscopic observation of a transition from a trivial to a topological phase in the engineered metamaterial. In contrast to prior work, our approach offers precise and independent in-situ tuning of the coupling parameters. Finally, we discuss the robustness of the topological edge state against various disorder realizations. Our results supplement efforts towards gate-controlled superconducting electronics and large controllable bosonic lattices to enable quantum simulations.
\end{abstract}

\maketitle

Metamaterials are engineered structures of simple constituents, which exhibit functionalities that go beyond those of the individual building blocks. For quantum systems, this transferable concept of metameterials can be used to engineer materials that host topological states of matter which remain robust against imperfections and are therefore suited for quantum computing and simulation~\cite{Kitaev2009topological, Sau2012}.
Recently, there has been great interest in topological metamaterials that resembles the Su-Schrieffer-Heeger (SSH) model~\cite{Su1979, Hasan2010, Asboth2016, Batra2020, Delnour2023}. The SSH model predicts a phase transition of a finite, 2N-site chain with alternating coupling strength between a band insulator and a topological insulator with localized edge states. The SSH model finds applications in the entanglement stabilization of quantum states~\cite{Jin2023}, the long-range interaction of qubits~\cite{Mei2018, Kim2021, Vega2021, Zheng2022} and the study of non-hermitian light-matter interaction~\cite{Galeano2021, Zelenayova2023, Io2023}. 

Due to the generality of the concept, topological metamaterials can be realized in various material platforms and efforts materialized in spin qubits~\cite{Kiczynski2022}, Rydberg atoms~\cite{DeLeseleuc2019}, adatoms~\cite{Pham2022} and integrated photonics~\cite{Ren22}. The idea has also been taken to classical electrical circuits made out of surface mount components on printed circuit boards~\cite{Liu2022, Iizuka2023, Zhou2023}. 
Here, we will focus on superconducting quantum circuits, which has emerged as one of the leading platforms for quantum simulation and computation~\cite{Blais2021}. Thanks to the mature fabrication and design techniques of superconducting circuits, the reliable large scale implementation of superconducting qubits and resonators is feasible which led to a number of experiments with engineered photonic baths~\cite{John1990, Liu2017, Mirhosseini2018, PuertasMartinez2019, Bello2019, Brehm2021, Ferreira2021, Nie2020, Scigliuzzo2022, Dmytruk2022, Ke2023}, topological electrical circuits~\cite{Lee2018, Jouanny2024band}, and qubit-to-topological waveguide coupling~\cite{Zhang2023, Pakkiam2023, Delnour2023, Wang2023, Kim2021}. 

While all of the aforementioned efforts resemble static realizations of metamaterials, the engineering of topological metamaterials from superconducting circuits with tunable spectral bandgap and controllable interaction remains an open challenge due to the necessary integration of multiple tunable elements with low cross-talk. Those efforts would enable quantum simulation of novel states of matter~\cite{Zhang2023-I, Karamlou2022} or the development of new circuit components such as on-chip isolators~\cite{Beck2023} or long-range couplers. The required tunability could be provided via flux control~\cite{Palacios-Laloy2008} or current biased conductors~\cite{Naaman2016,Visser2015, Xu2019, Parker2022}. However, a challenge still persist since it has been shown that flux-crosstalk between SQUID loops significantly complicates the device tune-up \cite{Abrams2019, Dai2021, Barrett2023}. It is also expected that long range supercurrents arising from current biased conductors may lead to crosstalk between otherwise independent sites. 
To alleviate the need for advanced control strategies and eliminate this type of crosstalk between mesoscopic circuit elements, one could instead leverage the local control over the microscopic properties of circuit elements. Such source of tunability became available with the local electrostatic control of the supercurrents in hybrid superconducting semiconducting structures~\cite{deLange2015, Casparis2019, Hertel2022, Splitthoff2022, Phan2023} or other hybrid structures, which should enable the crosstalk-free integration of tunability. Based on this concept of gate-tunability, gate-tunable transmon qubits~\cite{deLange2015, Hertel2022}, bus resonators~\cite{Casparis2019} and parametric amplifiers~\cite{Phan2023, Splitthoff2024} have been realized in previous work. 

In this work, we present a topological metamaterial with gate-tunable coupling within each unit cell. Specifically, we implement a five-unit-cell bosonic SSH chain comprised of two resonators per unit-cell forming a one-dimensional lattice of ten superconducting resonators. By taking the idea of electrostatic control from the single element level, as in a gatemon device, to scale, we achieve in-situ, electrostatic control over several inductive intra-cell coupling elements using semiconductor nanowire junctions, which enables the spectroscopic observation of a trivial-to-topological phase transition in the engineered topological metamaterial.

\begin{figure}
    \centering
    \includegraphics{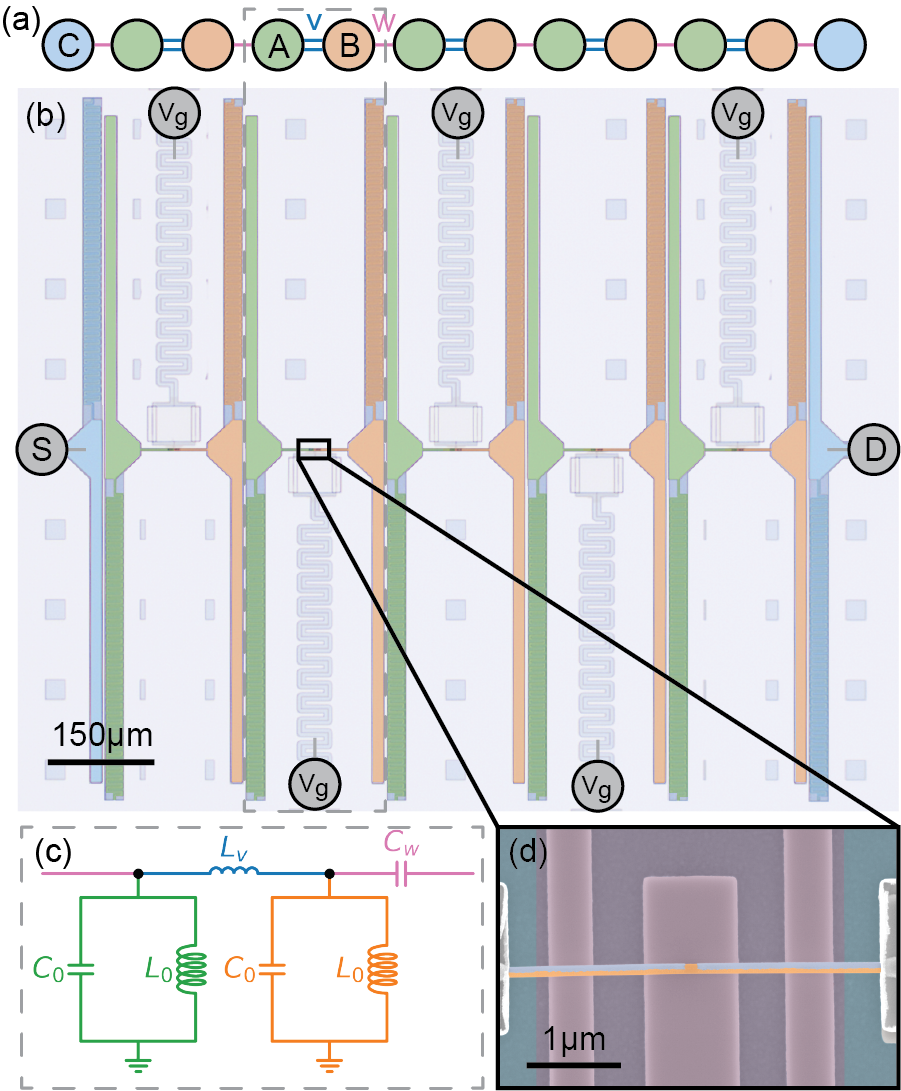}
    \caption{Superconducting resonator-based SSH chain implementation. 
    (a) Illustration of the SSH tight-binding model with two-site unit cell A,B, intra-cell coupling $v$ and inter-cell coupling $w$. The finite chain is terminated to coupling sites C. The gray box highlights one unit cell. 
    (b) False-colored microscope image of the five unit cell SSH chain composed of lumped element resonators with alternating variable inductive coupling and static capacitive coupling. The chain is measured from port S to port D in transmission. The inductive coupling, mediated by proximitized nanowire Josephson junctions, is controlled via five independence voltage gates. One unit cell is highlighted within the dashed box as in panel (a). 
    (c) Equivalent circuit of a single unit cell in (b). The individual circuit elements are discussed in the main text.
    (d) False-colored micrograph images of a single gate controlled proximitized nanowire Josephson junction. Orange: InAs semiconductor, blue: Al thin film, purple: SiN gate dielectric on NbTiN gate electrodes, turquoise: Si substrate, grey: NbTiN contacts. }
    \label{fig:Fig1_concept}
\end{figure}

We implement the five unit-cell tight-binding SSH model, as shown in Fig.~\ref{fig:Fig1_concept}a, using a chain of 10 lumped-element, high-kinetic inductance, superconducting resonators with alternating variable inductive intra-cell coupling $v$ and static capacitive inter-cell coupling $w$, see Fig.~\ref{fig:Fig1_concept}b. One unit-cell, see Fig.~\ref{fig:Fig1_concept}(c), is comprised of two resonators with intra-cell coupling $v$.
The capacitive inter-cell coupling is set by the geometry of the neighboring resonators between unit cells. 
The inductive intra-cell coupling is realized by five gate-tunable Josephson junctions formed in five proximitized semiconducting nanowires, which intra-connect the two resonators of a unit cell. 

The resonators are made from a high kinetic inductance film to reduce their footprints and avoid spurious modes. Within a unit cell, the resonators are spatially separated to minimize the residual capacitive intra-cell coupling and are arranged in a mirrored configuration to suppress the residual capacitive second nearest neighbor coupling and residual mutual inductances between neighbouring sites.
The chain is terminated to either side by resonator-like coupling sites with an open inductor to maintain a coupling strength $w$ on both ends to preserve the chiral symmetry. These coupling sites themselves connect directly to the input port S and the output port D, which enable transmission measurements. 

The equivalent circuit model of a single unit cell is shown in Fig.~\ref{fig:Fig1_concept}(c) displaying the four relevant circuit parameters: the resonator capacitance $C_0$, the resonator inductance $L_0$, the capacitive inter-cell coupling $C_w$, and the inductive intra-cell coupling $L_v$ on the two sub-lattice sites A and B. The relation between $v$ and $L_v$ and between $w$ and $C_w$ is discussed in the Supplementary Material Sec.~\cite{supp}. In this approximation we neglect the residual capacitive intra-cell $C_v$ and the residual capacitive second nearest neighbor coupling $C_{SNN}$.

As mentioned, the inductive coupling are realized using nanowire Josephson junctions as shown in the zoomed in picture in Fig.~\ref{fig:Fig1_concept}(d). The gate voltages are applied via Chebyshev-filtered gate lines [see Fig.\ref{fig:Fig1_concept}(b)] and control the Andreev bound states carrying the super-current between the superconducting leads of the nanowire junctions separately by changing the chemical potential which in turn set the respective Josephson energies $E_J$. Eventually, the coupling inductance is then given by the Josephson inductance $L_v(V_g)=\phi_0(2\pi I_c(V_g))^{-1}=\phi_0^2(E_J(V_g))^{-1}$, where $I_c$ is the gate-tunable critical current, $E_J$ the gate-tunable Josephson energy and $\phi_0=h/2e$ the magnetic flux quantum. The inductance approaches infinite when the nanowire Josephson junction gated below the pinched-off voltage $V_g< V_p$, and it is finite on the order of a few nano-Henry otherwise when charge carriers are accumulated in the junction region. Physically, the junction is comprised of thin Aluminum film on two facets of the InAs nanowire which has been selectively etched to form a \SI{110}{nm} long Josephson junction. The junction directly connects to resonators on either side via NbTiN contacts. The voltage gate is surrounded by two narrow ground lines to ensure equal ground potentials on the upper and lower half of the SSH chain.  

\begin{figure}
    \centering
    \includegraphics{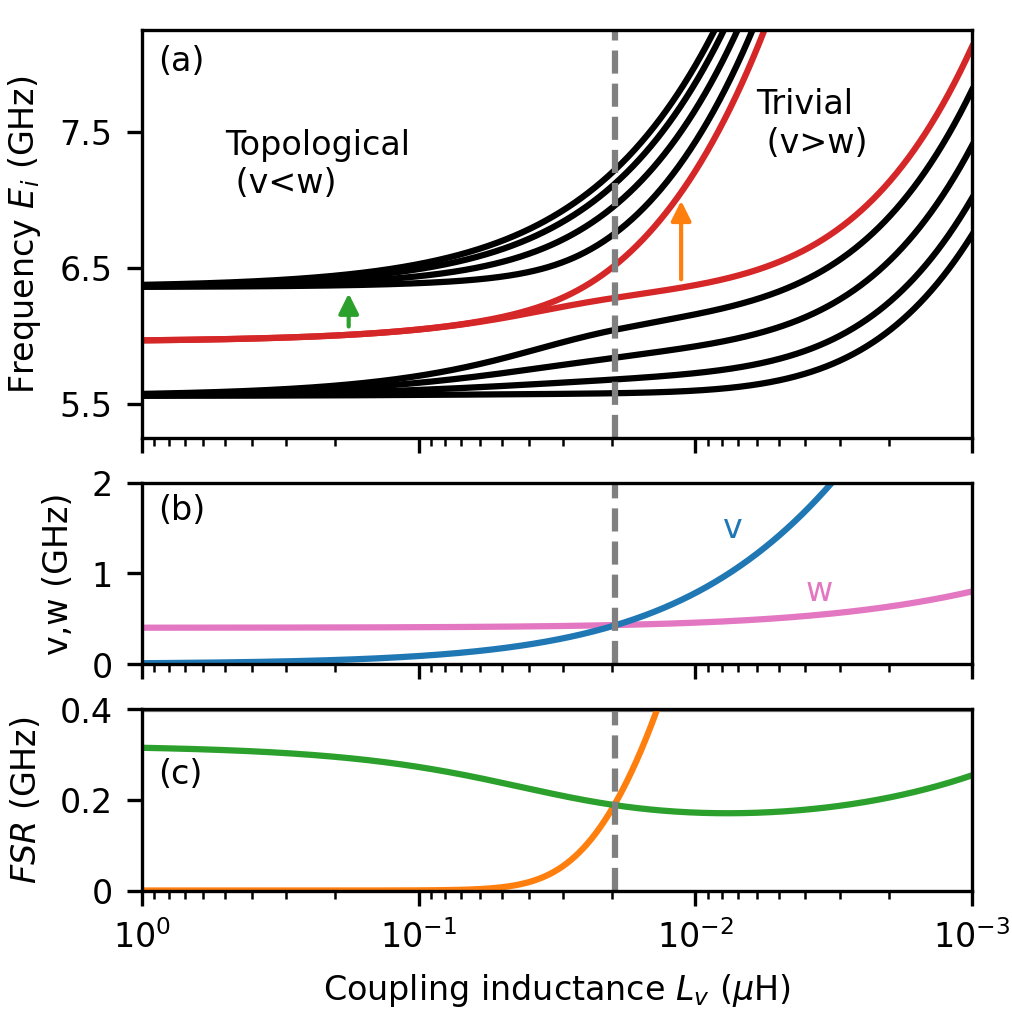}
    \caption{Simulated spectrum versus the coupling inductance $L_v$ of a bosonic SSH chain with the circuit parameters used in the experiment. 
    (a) Energy spectrum of the bulk modes (black) and the edge modes (red) versus the coupling inductance $L_v$. The dashed line indicates the phase transition point between the topological and the trivial phase where $v=w$. 
    (b) Coupling coefficients $v$ (blue), $w$ (pink) versus coupling inductance $L_v$.
    (c) Free spectral range $FSR$ between the edge states and the bulk (green, arrow in (a)) and between the two edge modes (orange, arrow in (a)).} 
    \label{fig:Fig2_SSHchain}
\end{figure}

Our circuit implementations shares the same tight-binding geometry as the SSH model with identical on-site potentials and alternating coupling strengths $v$ and $w$, as highlighted with a corresponding color code in Fig.~\ref{fig:Fig1_concept}(a,b). To further affirm the correspondence between the SSH model and our implementation we derive the real space Hamiltonian from the circuit Lagrangian (see Supplementary Material Sec.~\cite{supp}). 
The diagonalization of the real space Hamiltonian for realistic circuit parameters yields the eigenvalues for uniformly varied coupling strength $L_v$, see Fig.~\ref{fig:Fig2_SSHchain}. 

As shown numerically in Fig.~\ref{fig:Fig2_SSHchain}(a) for 10 circuit modes, in the case of large inductances, $L_v>\SI{22}{nH}$ corresponding to $v<w$, the spectrum exhibits two nearly degenerate mid-gap modes (red) centered around the eight bulk modes (black). Such mid-gap modes are the characteristic feature of a topological insulator state. Hence, we find the system to be in the topological insulator state if the nanowire junctions are pinched-off. As the coupling inductance decreases and the coupling ratio crosses over to $v>w$, see Fig.~\ref{fig:Fig2_SSHchain}(b), the mid-gap modes split and approach the bulk modes, which eventually leads to a fully gaped spectrum corresponding to the trivial insulator state. Particular for this implementation is the frequency shift of the entire spectrum for smaller coupling inductances as the change in $L_v$ also renormalizes the on-site potential, here the resonator frequency $\omega_r$. Consequently, also $w$ increases slightly following the increase in $\omega_r$ as seen in Fig.~\ref{fig:Fig2_SSHchain}(b). Note that the spectrum remains symmetric around the instantaneous resonator frequency $\omega_r$ (see Supplementary Material Sec.~\cite{supp}), which is characteristic for a system obeying chiral symmetry~\cite{Asboth2016}. 
To highlight the spectral evolution further, we present the free spectral range between one mid-gap mode and the nearest bulk mode as well as the separation between the two mid-gap modes in Fig.~\ref{fig:Fig2_SSHchain}(c). The system undergoes a phase transition between the topological and the trivial state as the order of the coupling strength $v$ and $w$ inverts. At the point where $v=w$, the free spectral ranges between the mid-gap modes and the nearest bulk modes are equal, see Fig.~\ref{fig:Fig2_SSHchain}(c), but remains finite due to the finite system size. We refer to this intermediate regime with approximately equal FSR as a normal state. 

Despite the presence of on-site and coupling strength disorder in this simulation, which represents the actual experimental circuit implementation, the spectrum is nearly indistinguishable from an ideal SSH chain, which shows the robustness of this implementation method. The only difference appears in the symmetry of the bulk modes (see Supplementary Material Sec.~\cite{supp}).

We experimentally resolve the energy spectrum of the 10 site SSH chain implementation via a transmission measurement at microwave frequencies through the chain from port S to port D for different gate voltages applied to nanowire junctions measured at the base temperature of a dilution refrigerator. In this measurement, eigenmodes of the SSH chain manifest as peaks in the transmission spectrum. Their linewidth is proportional to the respective wavefunction weight on the outer sites of the chain $|\psi_{0,2N}|^2$.

\begin{figure}
    \centering
    \includegraphics{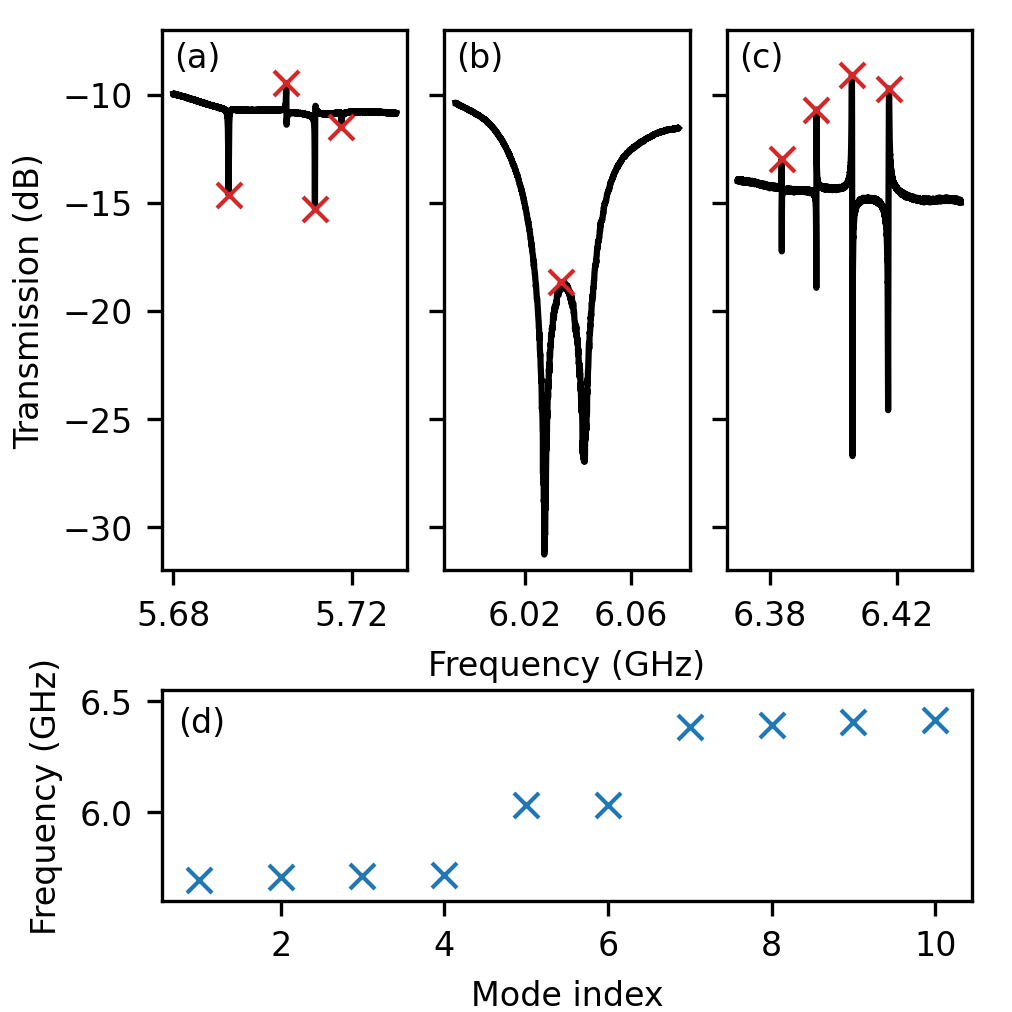}
    \caption{Parameter estimation of the SSH chain in the topological phase. (a-c) Transmission spectrum of the lower band, the gap and the upper band. The red crosses mark the 10 modes of the five unit cell SSH chain. (d) Eigenfrequency of the 10 modes in the topological phase sorted by mode index.}
    \label{fig:Fig3_parameterestimation}
\end{figure}

By pinching off the Josephson junctions, the intra-cell coupling $v$ is minimized and we measure the spectrum with a gate voltage of $V_g=\SI{-1}{V}$ on all gates. As shown in Fig.~\ref{fig:Fig3_parameterestimation}, we observe a transmission spectrum with 9 peaks corresponding to the eight bulk modes and the two quasi-degenerate edge modes occupying the lower band, the gap, and the upper band of the spectrum. The red crosses indicates the frequency points of the individual modes. As discussed, the linewidth of a mode depends on its coupling strength to the measurement ports. Hence, modes with more wavefunction weight on lattice sites close to the edge of the chain couple more strongly to the measurement ports and appear as broader modes in the spectrum. Following this argument, we assign the narrow modes in ranges \SIrange{5.68}{5.72}{GHz} and \SIrange{6.38}{6.42}{GHz} as bulk modes. The wider, central peak around \SI{6.04}{GHz} then corresponds to the quasi-degenerate mid-gap modes. The reduced transmission amplitude of the mid-gap modes compared to the maximal transmission results from the localization on either site of the chain. The asymmetric and distorted lineshapes of the SSH chain modes originate from Fano resonances of these relatively narrow modes with a broad spectral feature centered around \SI{6}{GHz} formed by the sample mount.  

The sorted eigenfrequencies versus mode index in Fig.~\ref{fig:Fig3_parameterestimation}(d) shows the gaped spectrum with quasi-degenerate zero energy modes, which is characteristic for the topological insulator state expected for this system. After the mode assignment, we can estimate the on-site energies and the coupling strength of each individual site along the SSH chain to obtain an estimate of the circuit parameters by fitting the full 10 dimensional Hamiltonian to the extracted eigenfrequencies (see Supplementary Material Sec.~\cite{supp}). The average values of the estimated circuit parameters $C_0$, $L_0$, $C_w$ and $L_v$ deviate from the design parameters by less than \SI{10}{\percent}. Moreover, we find a parameter disorder of less than \SI{1}{\percent} along the chain. The obtained parameters were used to generate the simulated data shown in Fig.~\ref{fig:Fig2_SSHchain}. 

\begin{figure*}
    \centering
    \includegraphics{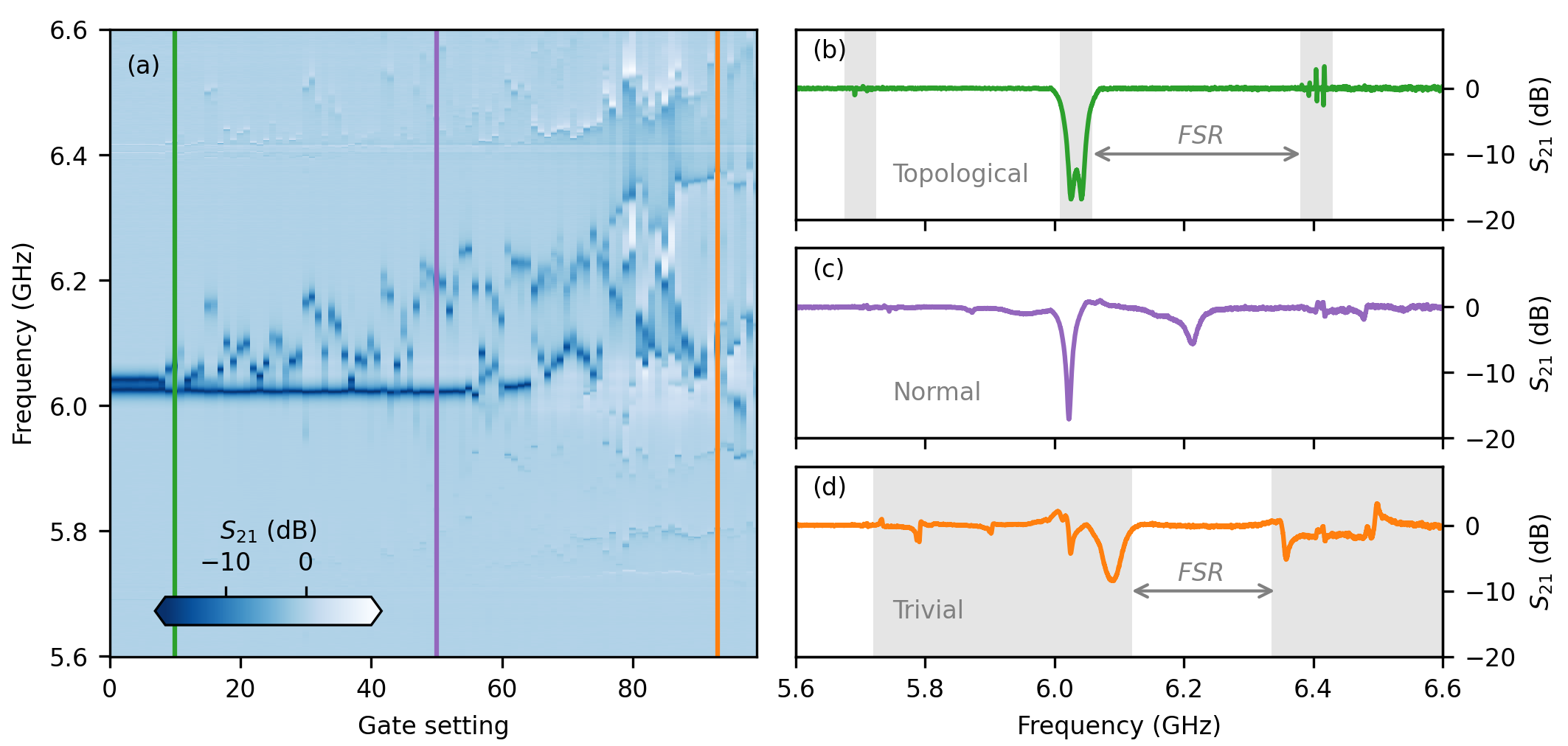}
    \caption{Gate-tunable phase transition. (a) SSH chain spectrum versus individual gate voltage $V_{g_i}$ aggregated in a joint gate setting sweep (more information in the main text). The shades indicate the normalized $S_{21}$ transmission through the chain. (b-d) SSH chain spectrum for a given gate setting indicated as colored line in (a). The greyed out regions in (b) and (d) highlight the width of the bands. The arrows indicate the largest free spectral range $FSR$ in these traces.}
    \label{fig:phasetransition}
\end{figure*}

Having established the circuit parameters in the topological regime, we now leverage the tunability per unit cell to observe the topological phase transition predicted for the SSH model. We first characterize the gate dependence of each individual nanowire junction. Specifically, we record the spectrum for a wide range of gate voltages per nanowire junction to identify the pinch-off voltage $V_{p_i}$ that suppresses the supercurrent across each junction and the open voltage $V_{o_i}$ that maximizes the supercurrent of the junction, see also Supplementary Material Sec.~\cite{supp}.  
Next, we measure the energy spectrum of the SSH chain along a synchronous evolution of the coupling inductances $L_v$ as we linear interpolate each junction between $V_{o_i}$ and $V_{p_i}$. We assume that each nanowire junction opens at a similar rate between $V_o$ and $V_p$ such that the joint gate control mimics the synchronous tuning of $L_v$ in all junctions. 
The resulting spectrum over \SI{1}{GHz} with a \SI{70}{kHz} resolution in the few photon regime versus a synchronous gate setting scan is shown in Fig.~\ref{fig:phasetransition}(a) after background correction (see Supplementary Material Sec.~ \cite{supp}). Three linecuts, colored relative to their respective gate setting in Fig.~\ref{fig:phasetransition}, exhibit three different regimes of the SSH chain and demonstrate that we can experimentally access each regime with in-situ tuning.

The spectral fan-out and overall increase of the eigenfrequencies of the SSH modes with increasing gate voltages, hence decreasing coupling inductances, qualitatively follows the simulation presented in Fig.~\ref{fig:Fig2_SSHchain}(a).
For the initial gate settings close to $V_p$ shown in Fig.~\ref{fig:phasetransition}(b), we obtain a gaped spectrum with mid-gap modes and a free spectral range of $FSR =\SI{350}{MHz}$, which we previously identified as topological insulator state. 
As the gate voltages increase towards $V_o$, the initially quasi-degenerate mid-gap modes with relatively wide linewidth and maximal prominence split and approach the lower and upper bundle of four modes. Eventually, close to $V_o$ the lines in the spectrum form two bundles of five modes each, are narrow in linewidth and of low prominence, see Fig.~\ref{fig:phasetransition}(d). We identify this single gaped spectrum with a free spectral range of $FSR =\SI{240}{MHz}$ as trivial insulator state. Note that the two highest mode in Fig.~\ref{fig:phasetransition}(d) exceed the measurement window. 

The phase transition of the SSH chain manifests in this experiment in the $FSR$ and the linewidth of the modes: (i) Starting from the topological insulator state, the $FSR$ between the mid-gap modes and the nearest bulk modes continuously closes while $FSR$ between the initial mid-gap modes widens. Around the phase transition the system passes through an intermediate, normal regime corresponding to the trace in Fig.~\ref{fig:phasetransition}(c), where the $FSR$ between the mid-gap modes and the $FSR$ to the nearest bulk modes is approximately equal. 
This change in $FSR$ is in qualitative agreement with the simulated trend shown in Fig.~\ref{fig:Fig2_SSHchain}(c). 
(ii) The narrower linewidth together with the larger transmission amplitude of the initial mid-gap modes indicate that these modes transition from being localized at the edges to becoming more delocalized as the inductors open. Moreover, as it can be seen from the comparison of Fig.~\ref{fig:phasetransition}(b) and Fig.~\ref{fig:phasetransition}(d), the bulk mode resonances widen from the topological to the trivial state (see Supplementary Material Sec.~\cite{supp}).

Beyond these two indicators of the phase transition, we also numerically calculate a topological invariant for the system, see Supplementary Material Sec.~\cite{supp}. Specifically, we calculate the real space winding number (RSWN)~\cite{mondragon2014topological, song2014aiii, rakovszky2017detecting,prodan2016bulk} and find a non-zero value in the topological regime ($v\approx 0$) while the RSWN approaches zero for more open junctions (increasing $v$). The finite size of the system prohibits a sharp transition between the two regimes, but the RSWN calculations verify that the limiting cases of pinched off and fully open represent two topologically distinct phases.

While the nonlinear frequency shift of the system is not directly related to the phase transition and kept small by choosing a low probe power, it serves as additional tool to verify that the nanowire junctions are open. While the system in the topological regime remains nearly unchanged versus changes in the probe power, the system frequencies lower drastically with power in the trivial regime due to added Kerr nonlinearity arising from the presence of the open Josephson junctions. 

A caveat, however, remains arising from the nature of nanowire Josephson junctions. As observed in earlier work on nanowire Josephson junctions~\cite{deLange2015, deMoor2018, Casparis2018, Pita-Vidal2023}, the functional dependence of the inductance on the individual gate voltage is highly non-monotonic and hysteretic over time, which prevents us from observing a smooth phase transition. The effect of disorder and disorder-induced phase transitions can further be studied, following ideas presented in Ref.~\cite{Mondragon2014, Perez-Gonzalez2019}. 

\emph{Conclusion.}
We have realized a macroscopic SSH chain out of a one-dimensional lattice of superconducting resonators, which inherits the tunability from the microscopic properties of nanowire Josephson junctions. We have characterized the system in the topological regime and mapped our bosonic implementation onto the original spin-less, single particle SSH Hamiltonian. Eventually, we leveraged the unit cell tunability and the engineered robustness against disorder to control the extended states along the SSH chain, which lead to the observation of the topological insulator phase transition in the macroscopic, gate-tunable resonator based SSH chain from the topological to the trivial insulator phase. 

Our experiment takes the idea of gate-tunable superconducting circuits~\cite{Xia2023} from a single gate-tunable element to scale and demonstrates that the implementation of several nanowire Josephson junctions in circuit QED experiments is possible. This result encourages further research on gate-tunable qubits, such as $\cos(2\phi)$ gatemon qubits~\cite{Larsen2020}, Andreev spin qubits~\cite{Pita-Vidal2023} or Kitaev chain qubits~\cite{Dvir2023, Pino2024}. Also the composition of two-dimensional lattices~\cite{Luo2023} of gate-tunable superconducting resonators seems feasible. 

However, our findings also suggest that further material development of the hybrid superconductor-semiconductor stack is required to enable a smooth and reproducible gate control and exploit the full potential of the microscopic properties of nanowire junctions in superconducting circuits. In future realization of this tunable system we intend to probe the on-site wavefunction with qubits as local probes via their affected decoherence and qubit frequency arising from an AC Start shift~\cite{Zaimi2019, Kim2021, Jin2023}. Going beyond the current implementation, one can study the nonlinear coupling~\cite{Hadad2018, Nair2022} between unit cells. Other work might focus on the long-range coupling of qubit to a chain or on implementing more complex systems like the Rice-Mele model~\cite{Rice1982} or the Kitaev model~\cite{Zhang2023-II}. The microwave transmission spectroscopy used in this work may also find application in the characterization of topological insulators and other material systems.

\section*{Acknowledgements} 
We thank Isidora Araya Day, Anton Akhmerov, Ana Silva and Dmitry Oriekhov for insightful discussions and Peter Krogstrup for the nanowire growth.
This research was co-funded by the Dutch Research Council (NWO) and by the Top consortia for Knowledge and Innovation (TKI) from the Dutch Ministry of Economic Affair. It is also part of the project Engineered Topological Quantum Networks (Project No.VI.Veni.212.278) of the research program NWO Talent Programme Veni Science domain 2021 which is financed by the Dutch Research Council (NWO). GJ acknowledges the research program “Materials for the Quantum Age” (QuMat) for financial support. This program (registration No. 024.005.006) is part of the Gravitation program financed by the Dutch Ministry of Education, Culture and Science (OCW).

\section*{Author contributions}
LJS, and CKA conceived the experiment with help from GJ and EG. LJS and MCB designed and acquired and analysed the data the sample. LJS fabricated the device. YL provided the proximitized nanowires. LJS, MCB and CKA wrote the manuscript with input from all other co-authors. CKA supervised the project.

\section*{Data availability}
The raw data and the analysis script underlying all figures in this manuscript are available online~\cite{Splitthoff2024-nwsshrepo}.

\bibliography{nwssh_ref}

\end{document}


\preprint{QuTech/AndersenLab}

\title{Supplementary Material for ``Gate-tunable phase transition in a bosonic Su-Schrieffer-Heeger chain''}

\author{Lukas Johannes Splitthoff$^{1,2}$}
\email{l.j.splitthoff@gmail.com}
\author{Miguel Carrera Belo$^{1,2}$}
\author{Guliuxin Jin$^{2}$}
\author{Yu Li$^{3}$}
\author{Eliska Greplova$^{2}$}
\author{Christian Kraglund Andersen$^{1,2}$}

\affiliation{$^1$QuTech, Delft University of Technology, Delft 2628 CJ, The Netherlands}
\affiliation{$^2$Kavli Institute for Nanoscience, Delft University of Technology, Delft 2628 CJ, The Netherlands}
\affiliation{$^3$Center for Quantum Devices, Niels Bohr Institute, University of Copenhagen, 2100 Copenhagen, Denmark}

\date{\today}

\maketitle


\section{SSH chain model}
Our implementation has been derived from the SSH model presented in Sec.~\ref{sec: condensed matter Hamiltonian}. We compare the model with the ideal implementation in Sec.~\ref{sec: LE, SSHchain, ideal} and discuss its symmetries and the localisation of the wavefunctions in Sec.~\ref{sec: SSHchain_ideal_sym} and Sec.~\ref{sec: SSHchain_ideal_loc}. 

\subsection{SSH Hamiltonian} \label{sec: condensed matter Hamiltonian}
The condensed matter system inspired tight-binding Hamiltonian of the SSH chain with on-site potential $\epsilon$, intra-cell hopping $v$ and inter-cell hopping $w$ yields a tri-diagonal real space Hamiltonian of dimension $2N$, where $N$ is the number of unit cells in the system. The basis of this real space Hamiltonian is formed by the sites along the chain. The Hamiltonian reads
\begin{equation}
    \begin{split}
        H_{CM} &= \sum_{n=1} ^{2N} \epsilon \ket{n}\bra{n} \\
        &+ \sum_n^{N} \big[v \ket{2n-1}\bra{2n}\\
        &+w\ket{2n}\bra{2n+1} + h.c. \big] \\
        & = \begin{pmatrix}
\epsilon & v &  & & & & \\
v & \epsilon & w & & & & \\
  & w & \epsilon & & & & \\
&   &   & \ddots & & & \\
 &  &  & &   \epsilon & w& \\
 &  &  & &  w& \epsilon & v\\
 &  &  & & & v & \epsilon\\
\end{pmatrix}
    \end{split}    
\end{equation}
The spectral properties of this system and its eigenstates are displayed in Fig.~\ref{fig:FigS2_SSHchain_comparison}. The key feature is the symmetrically gaped spectrum spanned by the bulk modes (black) with mid-gap modes in the topological insulator phase for $v<w$. The appearance of the symmetric spectrum is linked to the chiral or sublattice symmetry of the system, which can be tested by computing the anti-commutator 
\begin{equation}\label{eq:anticommute}
    \{\Gamma,H\} = \Gamma H + H\Gamma = 0
\end{equation}
with 
\begin{equation}
    \Gamma = \mathbf{I}_5 \otimes \sigma_z
\end{equation}
where the chiral symmetry operator $\Gamma$ is given by the outer product of $N$ dimensional identity matrix $\mathbf{I_5}$ and the Pauli-z matrix $\sigma_z$. For this tri-diagonal real space Hamiltonian described above the anti-commutator is indeed zero. The spectral gap never closes due to the finite size of the system, in contrast to the infinite limit in which the spectral gap is given by $|v-w|$. Instead, the transition point at $v=w$ is signaled by the equal free spectral range between the mid-gap modes and one of the mid-gap modes and the first bulk mode. Beyond the transition, the system enters a trivial phase where $v>w$ eventually leading to a full dimerization. The corresponding eigenstates of the mid-gap modes are localized around the edges of the chain, as exemplified for a specific configuration in Fig.~\ref{fig:FigS2_SSHchain_comparison}d, hence the name edge mode. An exponential fit to the wavefunction supported on either sublattice A or B yields the localization length $\xi$ in Fig.~\ref{fig:FigS2_SSHchain_comparison}b. As long as the edge modes are quasi-degenerate, the localization length extracted from the fit follows the theoretical limit $\log(w/v)^{-1}$, but the fit does not capture the divergence of the localization length well. For a finite system length, the inverse participation ratio appears to be a better measure for the localization of the wave function. 

\subsection{Lumped element, superconducting circuit Hamiltonian} \label{sec: LE, SSHchain, ideal}
\begin{figure*}
    \centering
    \includegraphics{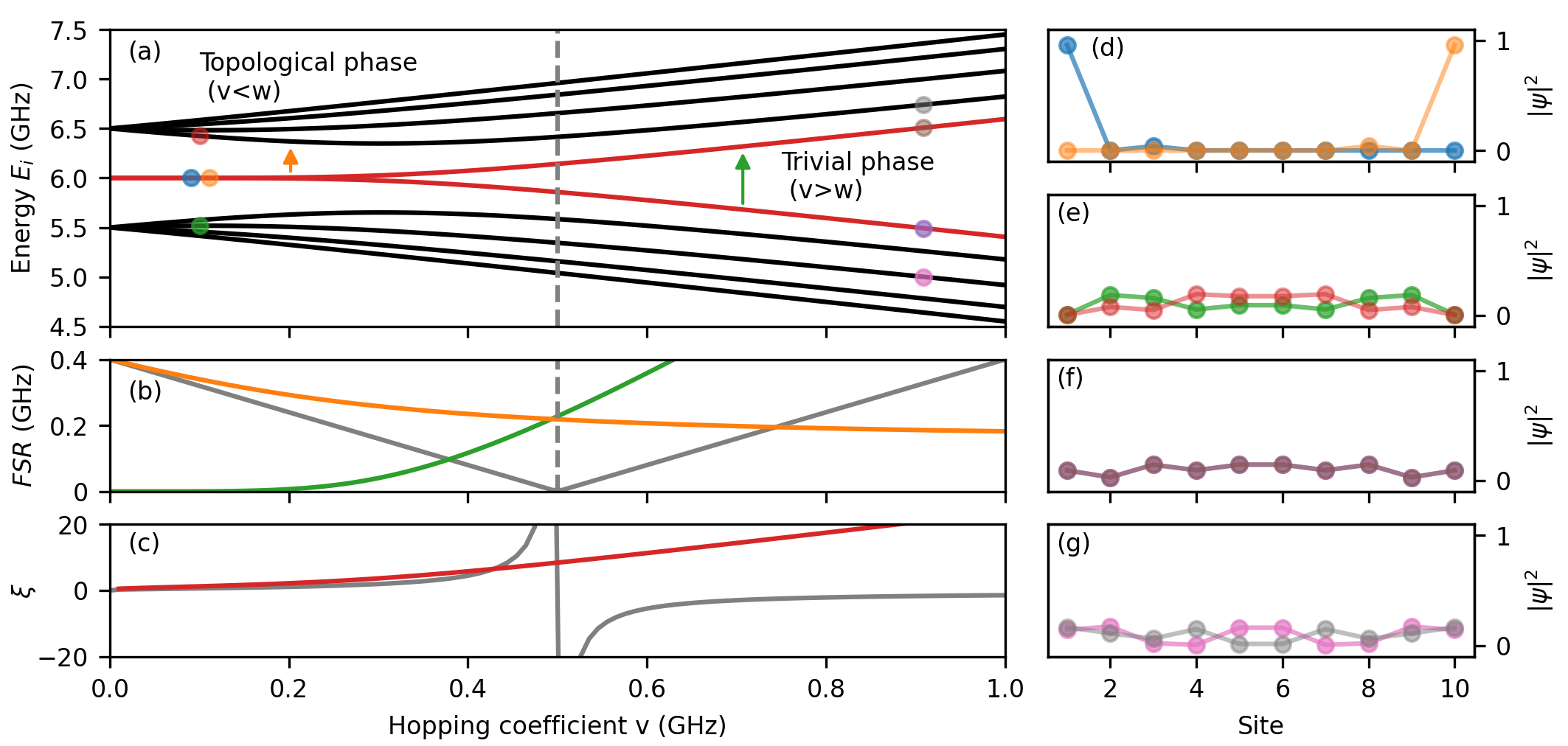}
    \caption{Simulated spectrum and states of a tight-binding SSH model with on-site potential $\epsilon=\SI{6.5}{GHz}$ and hopping coefficient $w=\SI{0.5}{GHz}$ 
    (a) Energy spectrum of the bulk modes (black) and the edge modes (red) versus the coupling strength $v$. 
    (b) Free spectral range $\Delta$ between the edge states and the bulk (blue, arrow in (a)) and between the two edge modes (orange, arrow in (a)) versus hopping strength $v$. The dashed line indicates the phase transition point between the topological and the trivial phase where $v=w$. The spectral gap $|v-w|$ (grey) only closes in the infinite length system 
    (c) Localization length $\xi$ extracted as exponential decay fit to the wave functions of the edge modes. The grey line shows the theoretical limit $\log(w/v)^{-1}$. (c) Energy spectrum of the bulk modes (black) and the edge modes (red) versus the coupling strength $v$. 
    (d-g) Real space wave function $|\psi|^2$ on the resonator lattice for the topological and the trivial phase of the bulk and the edge states. The colors correspond to the markers in (a).}
    \label{fig:FigS2_SSHchain_comparison}
\end{figure*}

The design of the lumped element, superconducting circuit implementation of the SSH chain follows a simple guiding principle. We realize every site as lumped element resonator with frequency $\omega$ and the coupling terms as capacitors $C_w$ or inductors $L_v$. We then derive the circuit Hamiltonian from a classical circuit Lagrangian formalism as part of a standard circuit quantization~\cite{Vool2017}. The resulting Hamiltonian $H_{LC}$ indeed takes the form of the tight-binding Hamiltonian
\begin{equation}
    \begin{split}
        H_{LC} &= \sum_{n=1} ^{2N} \hbar \omega \ket{n}\bra{n} \\
        &+ \hbar \omega \sum_n^{N} \bigg[\frac{L_T}{2L_v}\ket{2n-1}\bra{2n}\\
        &+\frac{C_w}{2C_T}\ket{2n}\bra{2n+1} + h.c. \bigg] \\
        & = \hbar \omega \begin{pmatrix}
1 & \frac{L_T}{2L_v} &  & & & & \\
\frac{L_T}{2L_v} & 1 & \frac{C_w}{2C_T} & & & & \\
  & \frac{C_w}{2C_T} & 1 & & & & \\
&   &   & \ddots & & & \\
 &  &  & &   1 & \frac{C_w}{2C_T}& \\
 &  &  & &  \frac{C_w}{2C_T}& 1 & \frac{L_T}{2L_v}\\
 &  &  & & & \frac{L_T}{2L_v} & 1\\
\end{pmatrix}
    \end{split}
\end{equation}
Based on the correspondence between the two systems, we can compare the coefficients in the Hamiltonian and formalize the mapping between the on-site energy $\epsilon$ and the resonator frequency $\omega$, between the coupling strength $v$ and the Josephson inductance $L_v$, and between the the coupling strength $w$ and the capacitive coupling $C_w$
\begin{align}
    \begin{split}
        \epsilon &= \hbar \omega\\
        v &= \frac{\hbar \omega}{2} \frac{L_T}{L_v}\\
        w &= \frac{\hbar \omega}{2} \frac{C_w}{C_T}
    \end{split}    
\end{align}
It should be noted that $\omega = \omega(L_v)$ which leads to a small increase in $w$, but a significant shift in the on-site potential $\epsilon$ causing the upwards shift of the spectrum as $L_v$ decreases.
For further visual comparison we also show the spectral properties and the eigenstates of the ideal SSH chain implementation in Fig.~\ref{fig:FigS2_SSHchain_comparison}.

\subsection{Topological properties of SSH implementation}
\label{sec:topoprops}
\subsubsection{Real space winding number}
Topological invariants characterize the global properties of topological insulators. In case of the SSH model, the winding number is used as the topological invariant, which takes a phase-dependent binary value: 0 in the trivial insulator phase and 1 in the topological insulator phase~\cite{ryu2010topological, asboth2016short}. 
In the finite-size SSH chain, the periodicity is absent, thus making momentum not a good quantum number and the winding number becomes an ill-defined invariant. Alternatively, the co-variant real space winding number (RSWN) serves as a good approximation for the momentum space winding number given the large enough system size, and the RSWN remains quantized in the presence of disorder~\cite{mondragon2014topological, song2014aiii, rakovszky2017detecting,prodan2016bulk}. In the following, we review the calculation of real space winding number. 

\begin{figure}
    \centering
    \includegraphics{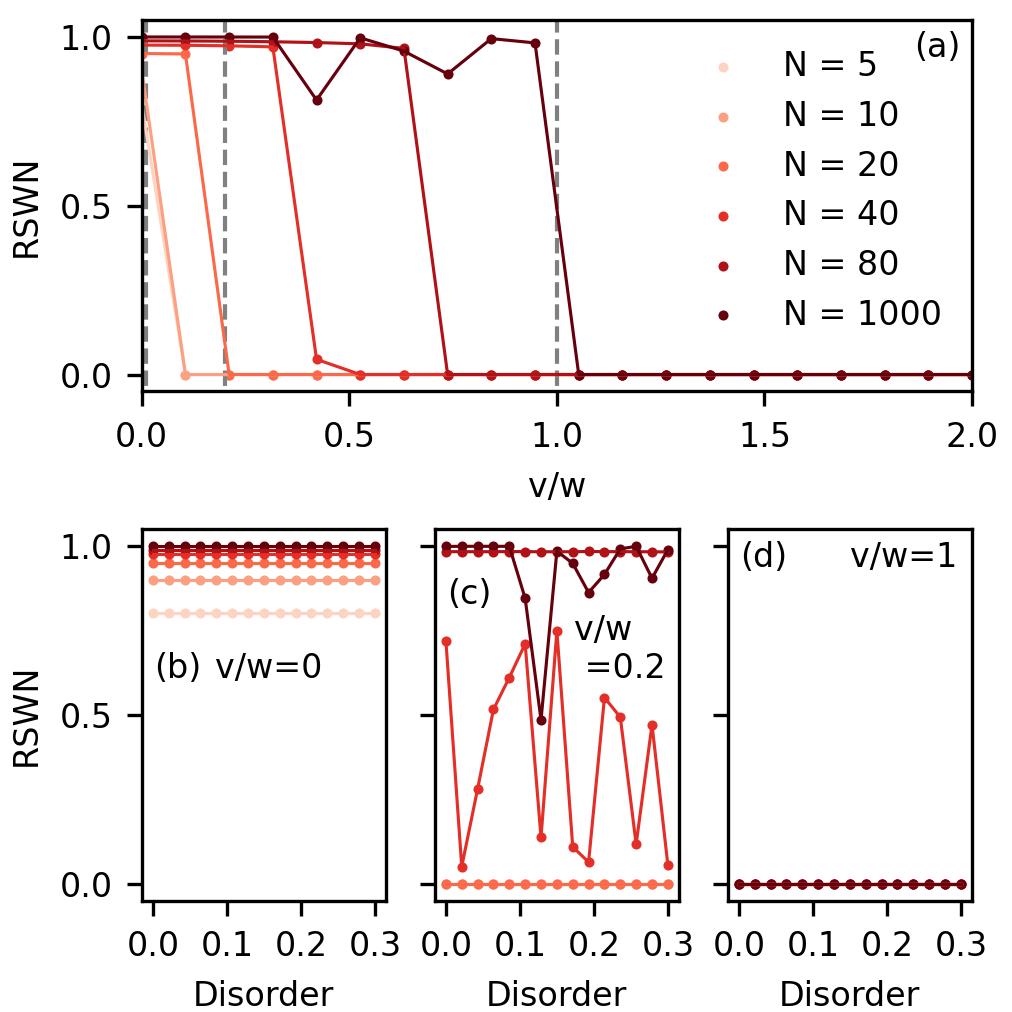}
    \caption{Real space winding number. (a) RSWN versus coupling strength ratio. The dashed vertical lines correspond to the $v/w$ ratio used in the disorder simulations shown in (b-d). (b-d) RSWN versus disorder strength for three different coupling strength.}
    \label{fig:realspacewindingnumber}
\end{figure}

For a 1D chiral symmetric system, recall the anti-commutation relation between the Hamiltonian $H$ and the chiral operator $\Gamma$ in Eq.~\eqref{eq:anticommute}. This relation results in the fact that the Bloch Hamiltonian $H(k)$ takes the off diagonal form~\cite{schnyder2009lattice}
\begin{equation}
    H(k) =
    \begin{pmatrix}
    0 & h(k)\\
    h(k)^{\dagger} & 0 \\
    \end{pmatrix}.
\end{equation}
The winding number is given by~\cite{schnyder2009lattice}
\begin{equation}\label{eq:kspacewinding}
    \nu = \frac{1}{2\pi i } \int_{BZ} \text{Tr} \{ [h(k)^{-1} \partial_{k} h(k)] \} \in \mathbb{Z}
\end{equation}
Eq.~\eqref{eq:kspacewinding} has a direct covariant form in the real space~\cite{mondragon2014topological}.
Starting from the real space SSH Hamiltonian $H$, the homotopically equivalent flatband version of $H$ is given by
\begin{equation}\label{eq:flatbandhamiltonian}
    Q = \frac{H}{|H|} = 
    \begin{pmatrix}
        0 & Q_0 \\
        Q_0^{\dagger} & 0 \\
    \end{pmatrix},
\end{equation}
where the off-diagonal part $Q_0$ enters the calculation of real space winding number. The co-variant real-space form of the winding number can be obtained by applying the Bloch-Floquet transformation of Eq.~\eqref{eq:kspacewinding} written in terms of flatband Hamiltonian~\cite{mondragon2014topological}
\begin{equation}\label{eq:rswn}
    \nu = - \text{Tr}_{\text{volume}}\{ Q_0^{-1} [X, Q_0]\},
\end{equation}
where $X$ is the position operator and $\text{Tr}_{\text{volume}}$ denotes the trace per volume. 
In the case of SSH model, the off-diagonal term $Q_0$ can be obtained as the following. The position eigenstates are $|x,c \rangle$, where $x = 1, \ldots ,L$ denote the unit cells and $c = A, B$ are the sublattice degree of freedom.
The position operator therefore takes the form $X = \sum_{x\in \mathbb{Z}} \sum_{c=A,B} x |x,c\rangle \langle x,c|$.
In order to obtain $Q_0$, we define spectral projectors as 
\begin{equation}
\begin{split}
    P_A =&\sum_{x\in \mathbb{Z}} \sum_{c=A} | x,c\rangle \langle x,c|,\\ 
    P_B =&\sum_{x\in \mathbb{Z}} \sum_{c=B} | x,c\rangle \langle x,c| = \hat{I} - P_A.
\end{split}
\end{equation}
The chiral operator $\Gamma$ is given by 
$\Gamma = P_A- P_B$.
In parallel, we define the projectors onto the upper and lower half of the energy spectrum
\begin{equation}
\begin{split}
    P_{-} &= \sum_{E_n \leq 0} |n\rangle \langle n|, \\
    P_{+} &= \sum_{E_n \geq 0} |n\rangle \langle n| =\Gamma P_{-} \Gamma.
\end{split}
\end{equation}
Then the flatband Hamiltonian in Eq.~\eqref{eq:flatbandhamiltonian} can be obtained by $ Q = P_{+} - P_{-} $. As a chiral-symmetric operator, $Q$ satisfies the condition $Q = P_A Q P_B + P_B Q P_A$.
The off-diagonal terms in Eq.~\eqref{eq:flatbandhamiltonian} can be written in terms of the above defined operators as 
\begin{equation}
    Q_{0} = P_A Q P_B, \quad (Q_{0})^{-1} = P_{B} Q P_A.
\end{equation}
Finally, one can express real space winding number Eq.~\eqref{eq:rswn} as 
\begin{equation}
    \nu = - \frac{1}{L} \text{Tr} \left\{ P_{B} Q P_A \left[X, P_A Q P_B \right] \right\}.
\end{equation}
In our study, we compute the RSWN versus the coupling strength $v/w$ for various chain lengths $N$ and for $w=0.5$, see Fig.~\ref{fig:realspacewindingnumber} (a). On either end of the coupling strength range ($v\in(0,1)$), the system approaches the the topological and trivial state with RSWN=1 and RSWN=0, respectively, regardless of the chain length. In-between $v\in(0,1)$, the RSWN indicates a cross-over between the two different insulator states, but only in the thermodynamic limit $N \rightarrow \infty$ the RSWN correctly predicts the phase transition point as it approximates the k-space winding number. For small $N$, the RSWN does not capture the phase transition, which is spectroscopically still defined, due to finite size effects.
Additionally, we investigate the robustness of the topological state by examining the RSWN under varying levels of disorder in the coupling parameter $v$ and $w$ for three different intra-cell couplings $v=\{0.01, 0.2, 1\}$, see Fig.~\ref{fig:realspacewindingnumber} (b-d). 
While the RSWN does not aim at benchmarking the phase transition for our finite size system, we see that in the limiting cases we have distinct topological phases which providing valuable insights into the topological phase transition that we observe experimentally.

\subsubsection{Symmetries} \label{sec: SSHchain_ideal_sym}
\begin{figure}
    \centering
    \includegraphics{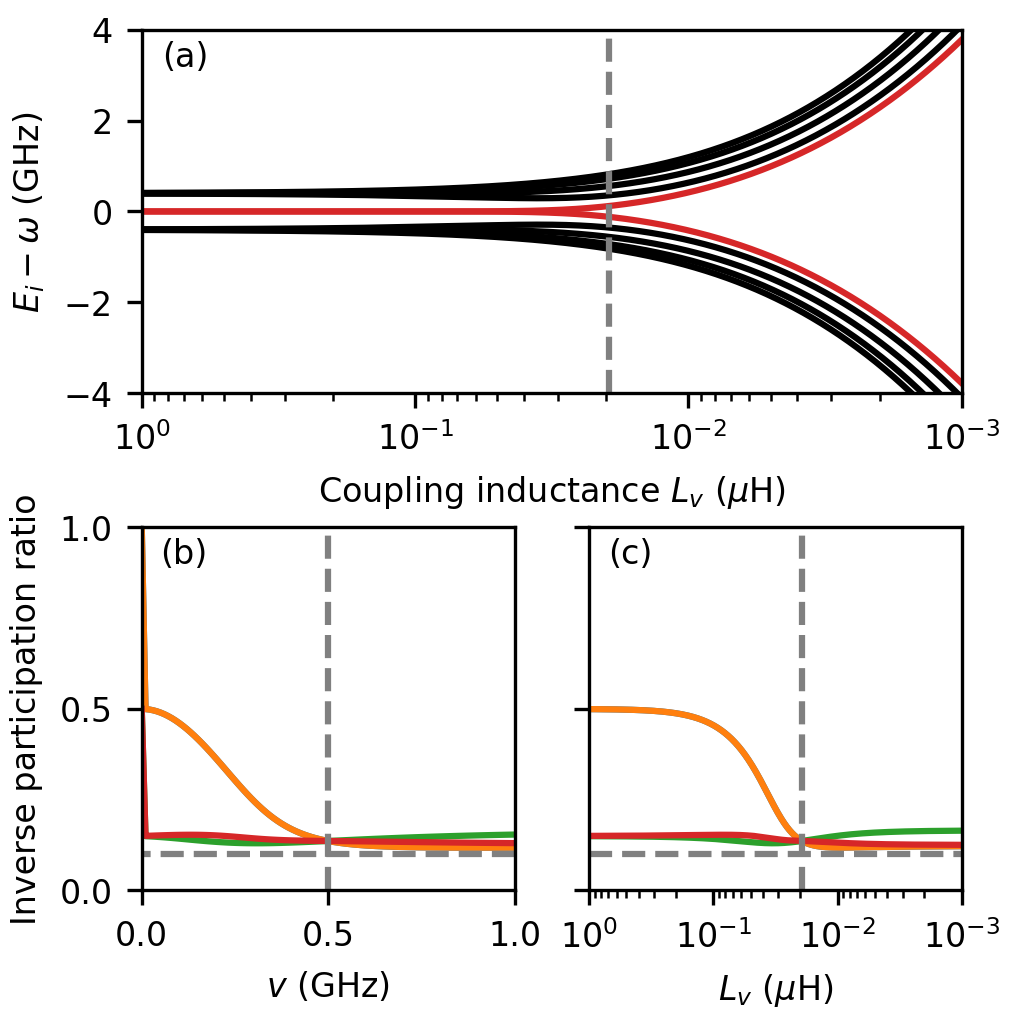}
    \caption{(a) Normalized simulated spectrum of the bosonic SSH chain with ideal the on-site and coupling energies as in Fig.~\ref{fig:FigS2_SSHchain_comparison}. (b,c) Inverse participation ratio for different states for the tight-binding model (b) and the ideal SSH chain implementation (c) versus the coupling strength $v$. The dashed line indicates the $1/2N$ limit. The colors correspond to the respective wave functions in Fig. \ref{fig:FigS2_SSHchain_comparison}.}
    \label{fig:FigB2_IPR_symmetry}
\end{figure}

To further strengthen the claim that the chiral symmetry persists in the ideal SSH chain implementation, we present the spectrum of Fig.~\ref{fig:FigS2_SSHchain_comparison}c normalized by the instantaneous resonator frequency $\omega(L_v)$ in Fig.~\ref{fig:FigB2_IPR_symmetry}(a). This normalized spectrum is symmetric around the on-site energy. Moreover, we can compute the anti-commutator and find that it vanishes $\{c,H_{LC}\}=0$.

\subsubsection{Localization} \label{sec: SSHchain_ideal_loc}
The localization of the wavefunctions of the edge states is a characteristic of the SSH chain. For finite chain lengths however, the localization length defined as exponential decay $\exp(-x/\xi)$ does not capture the delocalization of the edge modes as the system undergoes the phase transition, as seen in Fig.~\ref{fig:FigS2_SSHchain_comparison}(b). Instead, for finite chain length we can define an inverse participation ratio (IPR)~\cite{Scollon2020}, which measures the support of a wave function on a specific site and is defined as 
\begin{equation}
    IPR = \frac{\sum_x |\psi(x)|^4}{(\sum_x|\psi(x)|^2)^2}
\end{equation}
We observe that the IPR of the edge modes is finite and larger than for the bulk modes in the topological phase. As the system undergoes the phase transition, the IPR decreases and approaches the delocalization limit $1/2N=1/10$ of the bulk modes at $v=w$, see Fig.~\ref{fig:FigB2_IPR_symmetry}b,c. The IPR of the localized modes is of value one only in the fully dimerized case $v/w=0$, hence we observe a kink in the IPR in Fig.~\ref{fig:FigB2_IPR_symmetry}b close to $v=0$.

\section{Circuit parameter estimation}
\begin{figure}
    \centering
    \includegraphics{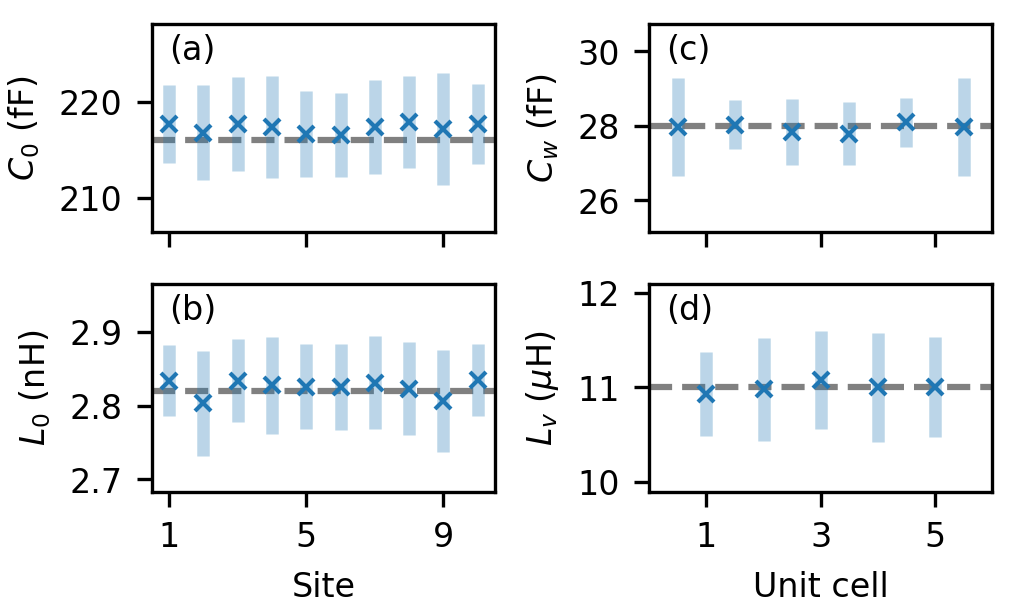}
    \caption{Extracted circuit parameters in the topological phase. (a,b) Resonator capacitance $C_0$ and inductance $L_0$ per site. (c,d) Coupling capacitance $C_w$ and coupling inductance $L_v$ per coupling site.}
    \label{fig:FigS5_extractedcircuitparameters}
\end{figure}

From the presence of the bulk and gap modes in the spectrum for a given $L_v$ configuration, we can extract the circuit parameters $C_0$, $L_0$, $C_w$ and $L_v$. First, we identify and list the eigenmodes of the system. Then, we fit the 10 dimensional Hamiltonian $H_{LC}$ to the list of eigenmodes while allowing for free variation of all circuit parameters along the chain. For completeness, we also take the two $C_w$ couplings to the coupling sites $C$ into account, which expands the $C_w$ list from four to six elements. Since the eigenfrequencies of the coupling sites are far detuned from the SSH spectrum, we do not consider them in the system Hamiltonian. The circuit parameters obtained from this optimization routine in the topological phase are shown in Fig.~\ref{fig:FigS5_extractedcircuitparameters}. We assume the design parameters as start parameters for the Nelder–Mead optimization. The variation of the obtained parameters is smaller than $\SI{1}{\percent}$. Finally, we input the obtained circuit parameters into the full Hamiltonian to compute the spectrum and the eigenstates to simulate the data in Fig.~2 and Fig.~3d. The agreement of this optimization with the measurement results is on the order of a few \SI{}{\kilo Hz}, which is also the spectral resolution of the measurement. 
This parameter estimation yields the static parameters $C_0$, $L_0$, and $C_w$. The optimization routine can then be repeated for different $L_v$ configurations to the trivial phase. 
\\While any on-site disorder and second-nearest neighbor interaction due to residual capacitive coupling breaks the chiral symmetry, strictly speaking, the mid-gap modes remain localized and can therefore be understood as chiral for practical purposes~\cite{Scollon2020}.  

\section{Single gate dependence}
\label{sec: single gate dependence}
\begin{figure*}
    \centering
    \includegraphics{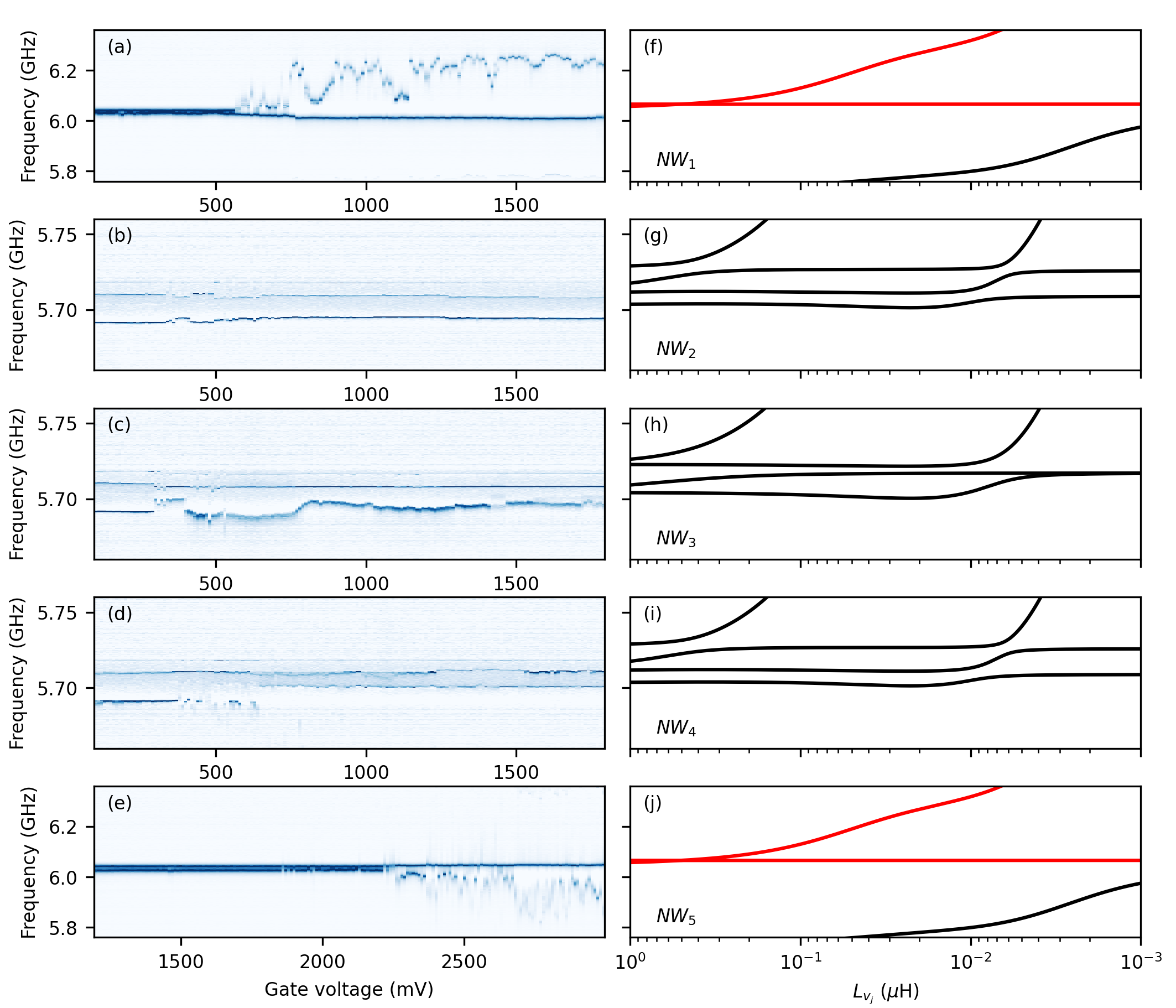}
    \caption{Single gate dependence. (a-e) Measured spectrum per gate while the other gates are set to $V_p$. Zoom-in on low band for bulk gates and zoom-in on gap for edge gates. The data is normalized on the background to enhance the visibility of the change. (f-j) Corresponding simulated spectrum.}
    \label{fig:singlegatedependence}
\end{figure*}

A transmission measurement through the five unit cell SSH chain allows for a site specific tune-up of every tunable coupling element. 
We run single gate scans for every nanowire Josephson junction over a wide gate voltage range, while we keep all other gates at the pinch-off voltage $V_p$, see Fig.~\ref{fig:singlegatedependence}(a-e). Hence, all scans begin deep in the topological regime. For the bulk gates $NW_i$ with $i\in [2,3,4]$ we zoom on the low band in the topological state, while we focus on the gap for the edge gates $NW_i$ with $i\in [1,5]$, as we expect the biggest change in frequency in those spectral ranges. 

We notice the non-monotonic single gate dependence of the modes in the respective test spectrum due to microscopic properties, most likely spurious quantum dots in the \SI{100}{nm} long junction. In fact, we observe discontinuities due to jumps in the microscopic properties of the junctions. Repetitions of these scans reveal that the slopes change over time. Hence, even these single gate dependencies cannot be used to calibrate for a joint gate scan across the phase transition point with identical nanowire inductances along the SSH chain. 

The measured single gate dependencies are in qualitative agreement with the simulated spectra on the right of Fig.~\ref{fig:singlegatedependence}. From the single gate dependencies we can extract the voltage $V_o$, which corresponds to the maximal supercurrent, hence minimal $L_J$ or, expressed in Josephson energy, maximal $E_J$. This voltage point seem robust enough and constant over time and over several gate scans. 

\section{Additional joint gate dependence}
\begin{figure}[t]
    \centering
    \includegraphics{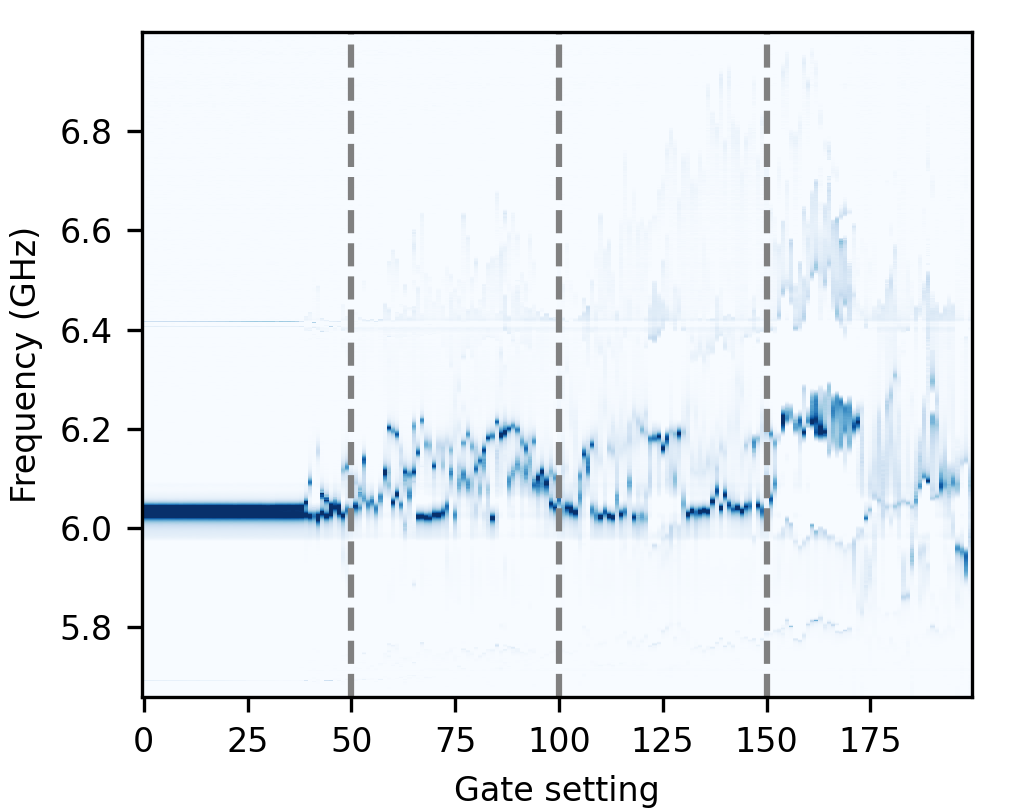}
    \caption{Additional joint gate dependence with $V_p$ and $V_o$ taken as minimal and maximal value of the single gate dependencies shown in Fig.~\ref{fig:singlegatedependence}. The dashed lines indicate the gate settings used for the power dependence shown in Fig.~\ref{fig:S10powerphasetrans}}
    \label{fig:2ndjointgatedepence}
\end{figure}

In Fig.~\ref{fig:2ndjointgatedepence}, we repeat the measurement shown in Fig.~4 over the gate settings $V_p$ and $V_o$ defined as minimal and maximal value of the single gate dependencies shown in Fig.~\ref{fig:singlegatedependence} at \SI{-30}{dBm} signal power. The splitting of the modes with increasing gate voltage, thus higher gate setting index, is in qualitative agreement with the equivalent scan from an earlier cooldown presented in Fig.~4, which uses similar gate settings $V_p=[400,150,430,400,1900]\SI{}{mV}$ and $V_o=[1800,1800,1800,1800,4000]\SI{}{mV}$ for the nanowire gates 1 to 5. 

\section{Power-induced phase transition} \label{app:powerphasetrans}
\begin{figure}
    \centering
    \includegraphics{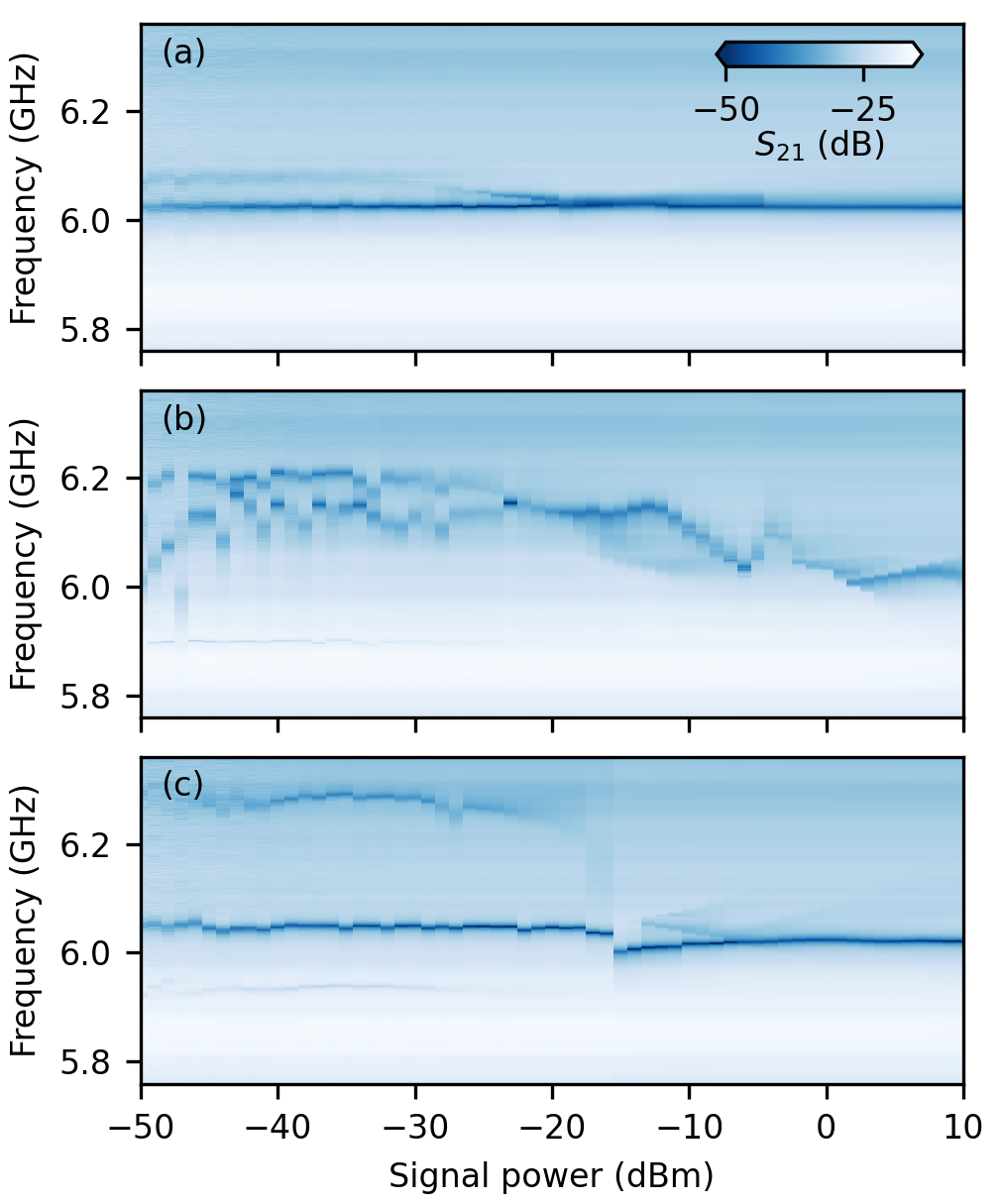}
    \caption{Power-induced phase transition. (a-c) Mid-gap spectrum versus signal power at the signal generator output for the three gate voltage settings [50, 100, 150] in Fig.~4.}
    \label{fig:S10powerphasetrans}
\end{figure}

Fig.~\ref{fig:S10powerphasetrans} shows the mid-gap spectrum versus signal power at the signal generator output for three gate voltage settings corresponding to the three gate settings [50, 100, 150] in Fig.~\ref{fig:2ndjointgatedepence}, (a) close to the topological regime, (b) close to the normal regime and (c) deep in the trivial regime. The dark lines on the light background indicate the mid-gap modes. 
We observe that the two modes are quasi-degenerate at high signal powers and the spectrum is nearly indistinguishable from the topological regime regardless of the exact gate setting. As we reduce the signal power, the two modes split and approach the low power spectrum shown in Fig.~\ref{fig:2ndjointgatedepence}. In the low power regime, the spectral evolution exhibits a gate dependence. For a gate set-point deep in the topological regime, the modes do not disperse with signal power (not shown here). We understand this power-induced phase transition as a result of the power-dependent change in the kinetic inductance of the proximitized nanowires. The inductance in these nanowires follows the equation 
\begin{align}
    L_{NW} = L_0(V_g) \bigg[ 1 + \bigg(\frac{I_s}{I_\ast(V_g)}\bigg)^2\bigg]
\end{align}
where $ L_0(V_g)$ is the gate-dependent low-power inductance, $I_s$ is the signal current through the chain, and $I_\ast(V_g)$ is the gate-dependent critical current. The signal power enters the equation via the signal current to which it is proportional $P_s ~ I_s^2$. Hence, an increasing signal power increases the inductance, which in turn transitions the SSH chain into the topological regime. The non-monotonic power dependence as well as the different transition points most likely arise from the different critical currents $I_\ast(V_g)$ per nanowire.
Consequently, in order to recover the SSH spectrum versus gate voltage, the measurements must be acquired with a signal power below \SI{-30}{dBm}.

\section{Fabrication} \label{app:devicefab}
We fabricate the SSH chain circuit and the gate lines from a \SI{40}{\nano \meter}-thick sputtered NbTiN film (kinetic inductance \SI{4}{\pico \henry \, \Box^{-1}}) on high resistivity n-doped Si. We pattern the NbTiN film using e-beam lithography and SF$_6$/O$_2$ based reactive ion etching. \SI{30}{\nano \meter}-thick plasma enhanced chemical vapour deposition (PECVD) SiN defined by a buffered oxide etch serves as bottom gate dielectric. We transfer the two-facet InAs/Al nanowire on top of the SiN bottom gate using a nano-manipulator. The InAs nanowires were grown by vapor–liquid–solid (VLS) growth with a diameter of \SI{110(5)}{\nano \meter}, and nominal thickness of the Al of \SI{6}{\nano \meter}~\cite{Krogstrup2015}. We selectively etch the \SI{110}{nm} long Josephson junctions into Al film. Then, we electrically contact the nanowires to the circuit via lift-off defined \SI{150}{\nano \meter}-thick sputtered NbTiN leads after prior Ar milling to minimize the contact resistance. 

\section{SSH chain environment} \label{sec:NWSSHbackground}
\begin{figure}
    \centering
    \includegraphics{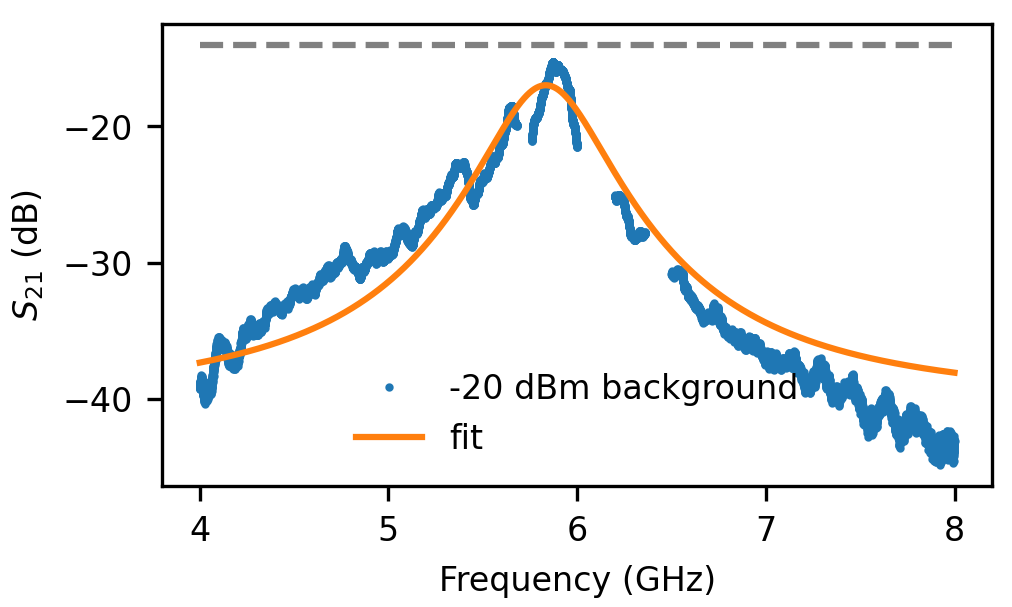}
    \caption{Box mode. $S_{21}$ transmission spectrum over the accessible frequency range $4-\SI{8}{GHz}$ with SSH spectrum removed (blue). A Lorentzian fit highlights the presence of a box mode (orange). The dashed grey line indicates the expected transmission given the input attenuation and output amplification.}
    \label{fig:FigS7_boxmode}
\end{figure}

We measure the transmission spectrum of the SSH chain in the topological state over the accessible frequency range ($4-\SI{8}{GHz}$), see Fig.~\ref{fig:FigS7_boxmode}. We observe a peaked background transmission with a finer modulation and with a maximal transmission around \SI{6}{GHz}. We attribute the resonance-like peak at around $\SI{6}{GHz}$ to a spurious feature hosted by the PCB enclosure spanning between the two test ports S and D, which acts as a band pass filter in parallel to the SSH chain, which in turn affects the overall transmission. We also observe that the transmission spectrum recorded in several experimental runs and on different samples, but within the same enclosure differs from the expected transmission spectrum. We highlight the resemblance of the transmission spectrum with a broad band resonator with a fit to a Lorentzian (orange). The finer modulation of the transmission probably arises from impedance mismatches along the lines connecting to the measurement electronics. To simplify the data analysis in the presence of the broad spectral feature, we obtain a linearly interpolated background transmission spectrum to which we normalize the measurement data to better identify the transmission peaks and dips of the SSH chain spectrum. 

\begin{figure}
    \centering
    \includegraphics{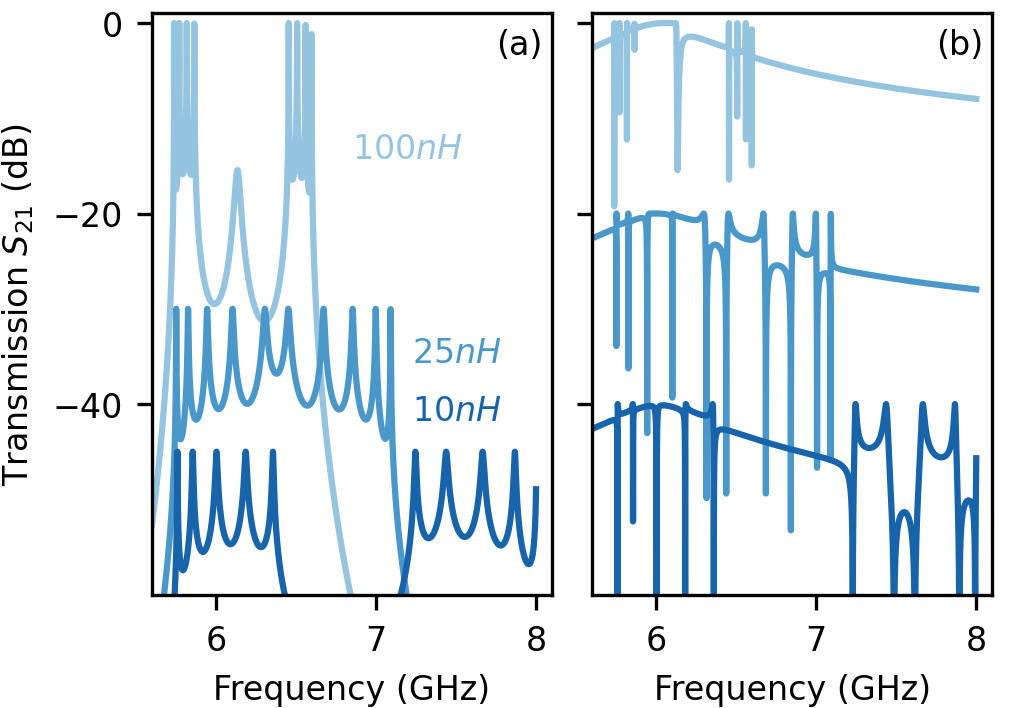}
    \caption{Lumped element simulation of the three regimes topological, normal, trivial for different tuning scenarios. (a) Expected, ideal SSH chain spectrum for a joint synchronous sweep of $L_v$. (b) Same device tuning as in (a), but accounting for the presence of a box mode modelled as transmission resonator in parallel to the SSH chain. Note that all spectra per panel are offset for better visibility.}
    \label{fig:Fig6_lumpedelementsimulation}
\end{figure}

While we cannot extract the exact effective inductance and capacitance leading to the spurious feature formed by the enclosure, we can approximate its effect in a lumped element simulation in which we assume a resonator in parallel to the SSH chain. The results of these lumped element simulations are presented in Fig.~\ref{fig:Fig6_lumpedelementsimulation}(a) for the ideal case and in Fig.~\ref{fig:Fig6_lumpedelementsimulation}(b) for the case in presence of a spurious resonance. In the ideal case, we expect two, or three well defined bands with in total 10 modes and otherwise suppressed background transmission. However, a spurious mode yields a higher overall transmission due its band pass feature and distorts the measured line shape of the modes due to is capacitive contribution on the rising edge and its inductive contribution on the falling edge. Consequently, the bulk modes appear as dips and peaks and the mid gap modes in the topological state appear as deep dip with a shallow peak in the center. 
The feature, which we modelled for simplicity as a single mode, does not affect the SSH chain, but it does complicate the interpretation of the overall transmission spectrum. To overcome the effect, a new sample enclose should be carefully designed~\cite{Huang2021}. Going beyond the current experiment, one could also intentionally coupled a SSH chain to a resonator as described in Ref.~\cite{Dmytruk2022}. 

\bibliography{nwssh_ref}